\documentclass[structabstract]{aa}  
\usepackage{graphicx}
\usepackage{subfigure}
\usepackage{longtable}
\usepackage{footmisc}
\usepackage{txfonts}
\usepackage{natbib}
\bibpunct{(}{)}{;}{a}{}{,}
\begin{document}
   \title{A Submillimetre Search for Cold Extended Debris Disks in the \mbox{{$\beta$} Pictoris} Moving Group\thanks{Based on observations with APEX, Llano Chajnantor, Chile (ESO programmes 079.F-9307(A) and 079.F-9308(A)).}}


   \author{R. Nilsson
          \inst{1}
          \and
           R. Liseau
          \inst{2}
          \and
           A. Brandeker
          \inst{1}
          \and
           G. Olofsson
          \inst{1}
          \and
           C. Risacher
          \inst{3}  
          \and
          M. Fridlund
          \inst{4}
          \and
          G. Pilbratt
          \inst{4}
          }
          
   \offprints{R. Nilsson}

   \institute{Department of Astronomy, Stockholm University, AlbaNova University Center, Roslagstullsbacken 21, SE-106 91 Stockholm, Sweden\\
              \email{ricky@astro.su.se, alexis@astro.su.se, olofsson@astro.su.se}
         \and
             Onsala Space Observatory, Chalmers University of Technology, SE-439 92 Onsala, Sweden\\
             \email{rene.liseau@chalmers.se}
         \and
         		 SRON, Postbus 800, 9700 AV Groningen, The Netherlands\\
         		 \email{crisache@sron.nl}
         \and
             ESA Astrophysics Missions Division, ESTEC, PO Box 299, 2200 AG Noordwijk, The Netherlands\\
             \email{malcolm.fridlund@esa.int, gpilbratt@rssd.esa.int}
         		 }

   \date{Received 9 March 2009 / 
   				Accepted [date]}

 
  \abstract
   {Previous observations with the \emph{Infrared Astronomical Satellite} and the \emph{Infrared Space Observatory}, and ongoing observations with \emph{Spitzer} and \emph{AKARI} have led to the discovery of over 200 debris disks, based on detected mid- and far infrared excess emission, indicating warm circumstellar dust. In order to constrain the properties of these systems, e.g., to more accurately determine the dust mass, temperature and radial extent, follow-up observations in the submillimetre wavelength region are needed.}
   {The $\beta$ Pictoris Moving Group is a nearby stellar association of young (${\sim}12\,$Myr) co-moving stars including the classical debris disk star $\beta$ Pictoris. Due to their proximity and youth they are excellent targets when searching for submillimetre emission from cold, extended, dust components produced by collisions in Kuiper-Belt-like disks. They also allow an age independent study of debris disk properties as a function of other stellar parameters.}
   {We observed 7 infrared-excess stars in the $\beta$ Pictoris Moving Group with the LABOCA bolometer array, operating at a central wavelength of 870\,{$\mu$}m at the 12-m submillimetre telescope APEX. The main emission at these wavelengths comes from large, cold dust grains, which constitute the main part of the total dust mass, and hence, for an optically thin case, make better estimates on the total dust mass than earlier infrared observations. Fitting the spectral energy distribution with combined optical and infrared photometry gives information on the temperature and radial extent of the disk.}
   {From our sample, $\beta$~Pic, HD\,181327, and HD\,172555 were detected with at least 3$\sigma$ certainty, while all others are below 2$\sigma$ and considered non-detections. The image of $\beta$~Pic shows an offset flux density peak located near the south-west extension of the disk, similar to the one previously found by SCUBA at the JCMT. We present SED fits for detected sources and give an upper limit on the dust mass for undetected ones.}
   {We find a mean fractional dust luminosity $\bar{f}_\mathrm{dust}=11{\cdot}10^{-4}$ at $t \approx 12\,$Myr, which together with recent data at 100\,Myr suggests an $f_\mathrm{dust} \propto t^{-\alpha}$ decline of the emitting dust, with ${\alpha}>0.8$.}

	 \titlerunning{A Submm Search for Cold Extended Debris Disks in the BPMG}
   \maketitle
%

\section{Introduction}
The first large and unbiased survey of the infrared (IR) sky, conducted by the \emph{Infrared Astronomical Satellite} (IRAS), revealed that approximately 15$\%$ of nearby main-sequence stars have an excess of 12--100\,$\mu$m emission, corresponding to a luminosity at least $2\times10^{-5}$ times higher than that expected from a pure stellar photosphere \citep{Backman1993,Backman1987,Aumann1984}. This indicates that these stars are surrounded by warm circumstellar dust, interpreted as originating in a disk of debris left over from planet formation, which is being heated by stellar optical and ultraviolet radiation, and re-radiating at mid- and far-IR wavelengths. The short orbital lifetime of small dust particles suggests that they are being continuously replenished by collisions of larger bodies \citep[see e.g.][]{Backman1995}. Deeper and more targeted observations by the \emph{Infrared Space Observatory} (ISO), \emph{Spitzer}, and \emph{AKARI}, have revised and extended this list of debris disk candidates to encompass over 200 stars \citep{Matthews2007,Rhee2007,DDDB}.

Although some of the most nearby debris disks have been imaged in the optical and IR, most are spatially unresolved and characterised solely on fits of the spectral energy distribution (SED) to stellar synthetic spectra and a handful of IR photometry data points. In order to better constrain the SED, and with this the temperature and radial extent of the disk, complementary submillimetre (submm) photometry is needed. Submm observations probe the Rayleigh-Jeans tail of the thermal radiation, where the measured integrated flux is roughly (sometimes with non-negligible correction) proportional to the temperature and the mass of the dust. Since the most efficient submm emitters are large and cool grains, which also dominate the mass of the disk \citep{Zuckerman2001}, observations with instruments like the Large APEX BOlometer CAmera \citep[LABOCA,][]{Siringo2007} on the Atacama Pathfinder EXperiment \citep[APEX,][]{Gusten2006} telescope (operating at 870\,{$\mu$}m) or SCUBA \citep{Holland1999} at the James Clerk Maxwell Telescope \citep[JCMT,][]{Prestage1996} (operating at 850\,{$\mu$}m) provide good estimators for the total dust mass (assuming an optically thin disk at these wavelengths).

Another advantage of observations in the submm regime is the ability to investigate cool dust created in very extended debris disks or belts located some hundreds of AU from the star (akin to the Solar system's Kuiper Belt), compared to the inner warm dust at 1--100~AU radius studied in IR surveys. The larger disk region probed by submm, compared to IR, makes it feasible to potentially resolve nearby disks, and morphologically determine outer disk radii, dynamical interaction with unseen planets, etc.~\citep[e.g.][]{Wyatt2008}, even though the angular resolution in the submm in general is lower than for shorter wavelengths.

Young, nearby stars, with a confirmed mid- and far-IR excess are prime targets for a submm search for cold extended disks. The $\beta$ Pictoris Moving Group \citep[BPMG,][]{Zuckerman2001a} is a nearby stellar association of 30 identified young ($\sim12$\,Myr) co-moving member stars \citep{Zuckerman2004,Barrado1999}, harbouring the perhaps most studied of all debris disk systems, namely $\beta$ Pictoris \citep{Smith1984}, and also e.g., AU~Mic. In this paper we present a study of 7 main-sequence IR excess stars in the BPMG, observed at 870\,$\mu$m with LABOCA at APEX in Chile. The stars were selected from the sample of currently known members of BPMG fulfilling the following criteria: (1) located at $\delta<-50\degr$, thus, southern stars that are not accessible to JCMT, but that will be observable with ALMA, and (2) previously detected IR excess. However, several of the stars previously observed at 850\,{$\mu$}m, e.g.~AT~Mic and AU~Mic, had to be excluded due to time constraints. By targeting members of the same moving group we ensure that any differences in dust properties and mass reflect intrinsic, age-independent, variations, since these stars are assumed to be coeval \citep{Mentuch2008}. This investigation is the precursor of a larger survey of 10--100~Myr old stellar associations, which will permit a statistical survey of disk properties during crucial time periods of debris disk evolution. The ultimate goal is to determine: the incidence of cold dust disks around main sequence stars; whether these disks can serve as indicators of planetary systems; the physical characteristics, e.g. chemical composition and grain size distribution, as a function of fundamental stellar parameters such as mass, metallicity and age; and the disk lifetimes.


\section{Observations and Data Reduction}
Since May 2007, the 12-m diameter submm telescope APEX, operated jointly by Onsala Space Observatory, the Max-Planck-Institut f{\"u}r Radioastronomie, and the European Southern Observatory at an altitude of 5100~m in the Chilean Andes, has been equipped with the LABOCA bolometer array. It operates at a central wavelength of 870\,{$\mu$}m (345\,GHz) and a bandwidth of 150\,{$\mu$}m (60\,GHz), covering a 11$\farcm$4 field-of-view with its array of 295 bolometers; each with a 18$\farcs$6 full-width at half-maximum (FWHM) beam.

Observations were performed between July 27 and August 5, 2007, employing a spiral pattern scan in order to recover fully sampled maps from the under-sampled bolometer array \citep{Siringo2007}. Each 7.5\,min long scan gives a raw map with uniform noise distribution for an area of about 4$\arcmin$ around the central position of the source. Targets and observation data are presented in Table \ref{table:1}. Additional calibration measurements to determine the pointing accuracy were made on Jupiter. Uranus and Neptune were observed for absolute flux calibration together with secondary calibrators B13134, G5.89 and IRAS16293, while several skydips in-between target observations yielded the correction for atmospheric opacity (with a zenith opacity ranging between 1.0 and 3.0).

\begin{table*}
\caption{Observed objects and integration time}
\label{table:1}
\centering
\begin{tabular}{l c c c c c}
\hline\hline
Target & R.A. & Dec. & Spectral Type & Distance & Integration time\\
 & (h m s) & ($\degr$ $\arcmin$ $\arcsec$) &  & (pc) & (h:min) \\
\hline
HD\,15115 & 02 26 16.24 & +06 17 33.2 & F2 & 45 & 1:45 \\

HD\,39060 ($\beta$~Pic) & 05 47 17.09 & -51 03 59.5 & A6V & 19.3 & 10:45 \\

HD\,164249 & 18 03 03.41 & -51 38 56.4 & F5V & 47 & 4:00 \\

HD\,172555 & 18 45 26.90 & -64 52 16.5 & A7V & 29 & 2:00 \\

HD\,181296 & 19 22 51.21 & -54 25 26.2 & A0Vn &  48 & 1:45 \\

HD\,181327 & 19 22 58.94 & -54 32 17.0 & F5/F6V & 51 & 2:08 \\

HD\,191089 & 20 09 05.21 & -26 13 26.5 & F5V & 54 & 1:23 \\
\hline
\end{tabular}
\end{table*}

%
%
%
%
%
%
%
%

For data reduction, we used mainly the Java-based software MiniCRUSH, a special version of CRUSH \citep[Comprehensive Reduction Utility for SHARC-2,][]{Kovacs2008} adapted for the APEX bolometers, while IDL and Matlab were used for further analysis and calculations. The reduction procedure in MiniCRUSH started with a correction for the atmospheric opacity calculated from observed sky temperature during the skydips, with cryostat temperature drifts taken into account. Each target scan was corrected with linearly interpolated opacity values from its nearest preceding and nearest subsequent skydip scan. Thereafter, the target scans were similarly flux calibrated from neighbouring calibrator measurements.
After these initial corrections the time-series signal of each bolometer and for each scan was reduced by incremental modelling, following the general reduction steps listed below:
\begin{enumerate}
	\item 1/f-filtering (over 30\,s scales for first iteration, then over 10 $\times$ crossing times)
	\item Corrections for correlated sky noise
	\item Normalisation with mean sky gain (flat-fielding) after flagging gains outside 0.1--10
	\item Channel weighting
	\item Despiking at 100, 30 and 10$\sigma$ after flagging channels in the array moving too fast or too slow during the spiral pattern scan, as well as dead and very noisy channels
	\item Corrections for correlated noise between bolometers sharing some parts of the electronics
	\item Normalization to mean bolometer gain (again after flagging gains outside 0.1--10)
	\item Noise whitening (in last iteration)
	\item Source mapping by discarding and blanking at 30, 10 and 3$\sigma$ (except last iteration), exposure clipping at 10\% peak integration, and scan weighting via median noise estimation.
\end{enumerate}	
This was iterated 5-6 times to attain a converged solution and recover a final map of our target \citep{Kovacs2008}.

After closer inspection of our calibration-target (point-source) scans we noticed non-negligible differences between peak fluxes and integrated fluxes hinting at a possibly de-focused beam, which should be taken into account when interpreting the spatial extent of any detected sources (see section \ref{sec:dis}). This should not have any effect on measured integrated fluxes, however, the calibration of data from submm imaging arrays is non-trivial since strong, point-source calibrators are not reduced in the same way as, e.g.~weak extended objects. The differences between the reduction procedures have to be quantified in order to accurately compare source and calibration images \citep[see][]{Kovacs2006}.


\section{Results}
Out of our sample of 7 main-sequence stars in the BPMG, two were clearly detected in the reduced 870\,{$\mu$}m maps, while one additional star showed what is most likely a source-associated submm emission, at a 3--4$\sigma$ level. Root-mean-square (RMS) noise weighted average maps of the three objects are presented in Fig.~\ref{fig:1all}a--c. The colour scale representing the flux level has been truncated at zero in order to give cleaner maps. Each image has been smoothed with a 9$\farcs$3-diameter circular Gaussian representing half the FWHM size of the beam, yielding an effective image resolution FWHM of 20$\farcs$8. The first solid contour corresponds to a 2$\sigma$ flux level with the following contours at increments of 1$\sigma$ for HD\,181327 and HD\,172555, and 2$\sigma$ for $\beta$~Pic, while the dotted contour outlines the 1$\sigma$ level. The position of the star as given in SIMBAD astronomical database \citep{Wenger2000} corresponds to $(\Delta\alpha,\Delta\delta)=(0\arcsec,0\arcsec)$. The total integrated flux density for $\beta$~Pic and HD\,181327 was found by fitting a Gaussian to a defined circular source region, with baseline level and gradient correction as measured outside the region. For the special case of HD\,172555 the integrated flux was measured within the rectangles plotted in Fig.~\ref{fig:1all}c. Fluxes and RMS-noise values $\sigma$ of the obtained images are presented in Table \ref{table:2}, together with the inferred dust temperature, power law index of the opacity law, dust mass, and fractional dust luminosity (see section \ref{sec:dis}). For undetected sources an upper ($3\sigma$) limit on the dust mass is given. HD\,15115 was detected at 850\,$\mu$m by \citet{Williams2006}, who calculated a dust mass of 3.9\,$M_\mathrm{Moon}$ using $\kappa_{850}=1.7$\,cm$^{2}$g$^{-1}$. Considering flux- and temperature errors, and our use of a slightly higher $\kappa$-value, this is consistent with our derived upper detection limit, and it probably just barely escaped detection in our observations.



\subsection{$\beta$~Pictoris}
The flux density peak of $\beta$~Pic (Fig.~\ref{fig:1all}a) is 46.7\,mJy/beam, centered on the position of the star, but with the source just barely resolved (with a FWHM of the fitted Gaussian equal to $24{\arcsec}.8$), causing the dust emission not to appear significantly elongated. Previous submm observations with SCUBA at JCMT found a marginal extension of the disk with a position angle (PA) of 32$\pm$4$\degr$ \citep{Holland1998}, which is similar to the PA of the circumstellar disk measured from IR and optical observations \citep[e.g. 30.7$\pm$4$\degr$ from][]{Kalas1995}. Also interesting is the offset (${\Delta}\alpha,{\Delta}\delta)=(-26{\arcsec},-22{\arcsec})$ south-west (SW) 13.1\,mJy/beam (5$\sigma$) flux-peak (labelled I in the figure) with a PA of 36$\pm$5$\degr$ lying slightly off the disk plane, at a position roughly consistent with the offset $(-21{\arcsec},-26{\arcsec})$ ``blob'' (position labelled A) found by \citet{Holland1998}. There is no significant flux at the location (labelled B) of the 1200\,$\mu$m peak found by \citet{Liseau2003}. The far SW 6$\sigma$ flux-peak (labelled II) is probably a background submm galaxy, but coincidentally has the same PA as our measured peak I. Both these features will be discussed further in section \ref{sec:dis_pic}.

\subsection{HD\,181327}
\begin{figure*}
     \centering
     \subfigure[]{
          \label{fig:1a}
          \includegraphics[width=.45\textwidth]{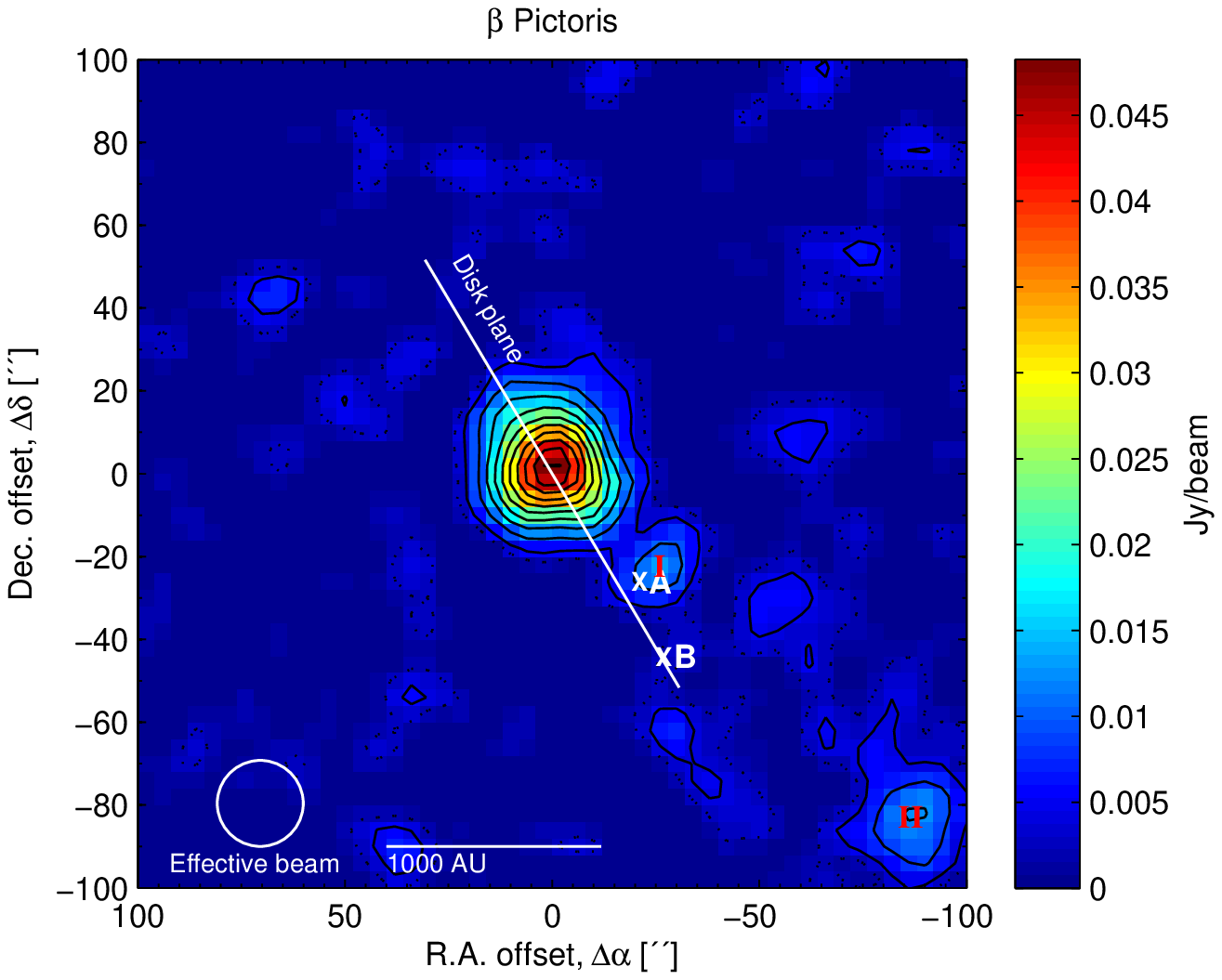}}
     \hspace{1mm}
     \subfigure[]{
          \label{fig:1b}
          \includegraphics[width=.45\textwidth]{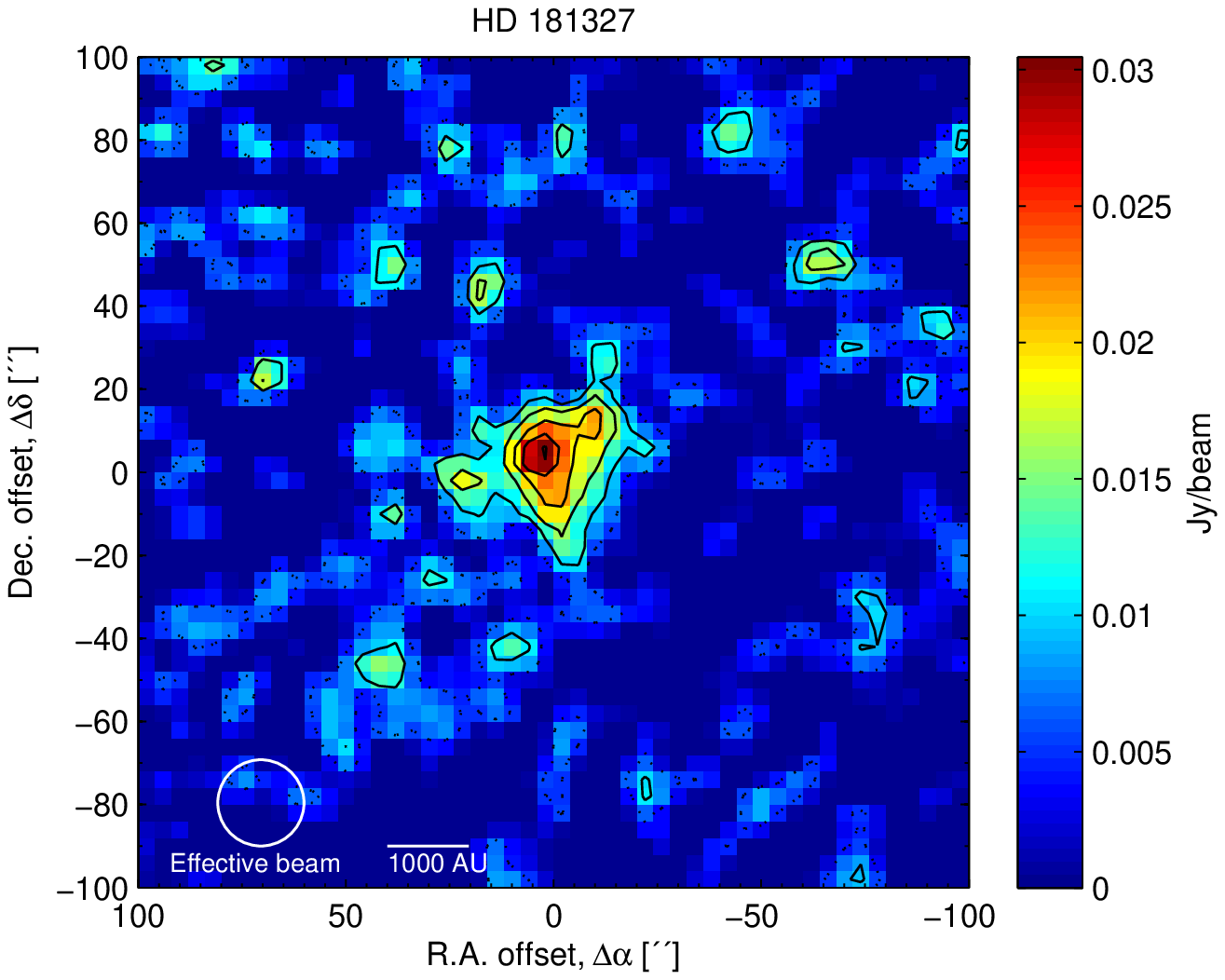}} \\
     \vspace{1mm}
     \subfigure[]{
           \label{fig:1c}
          \includegraphics[width=.45\textwidth]{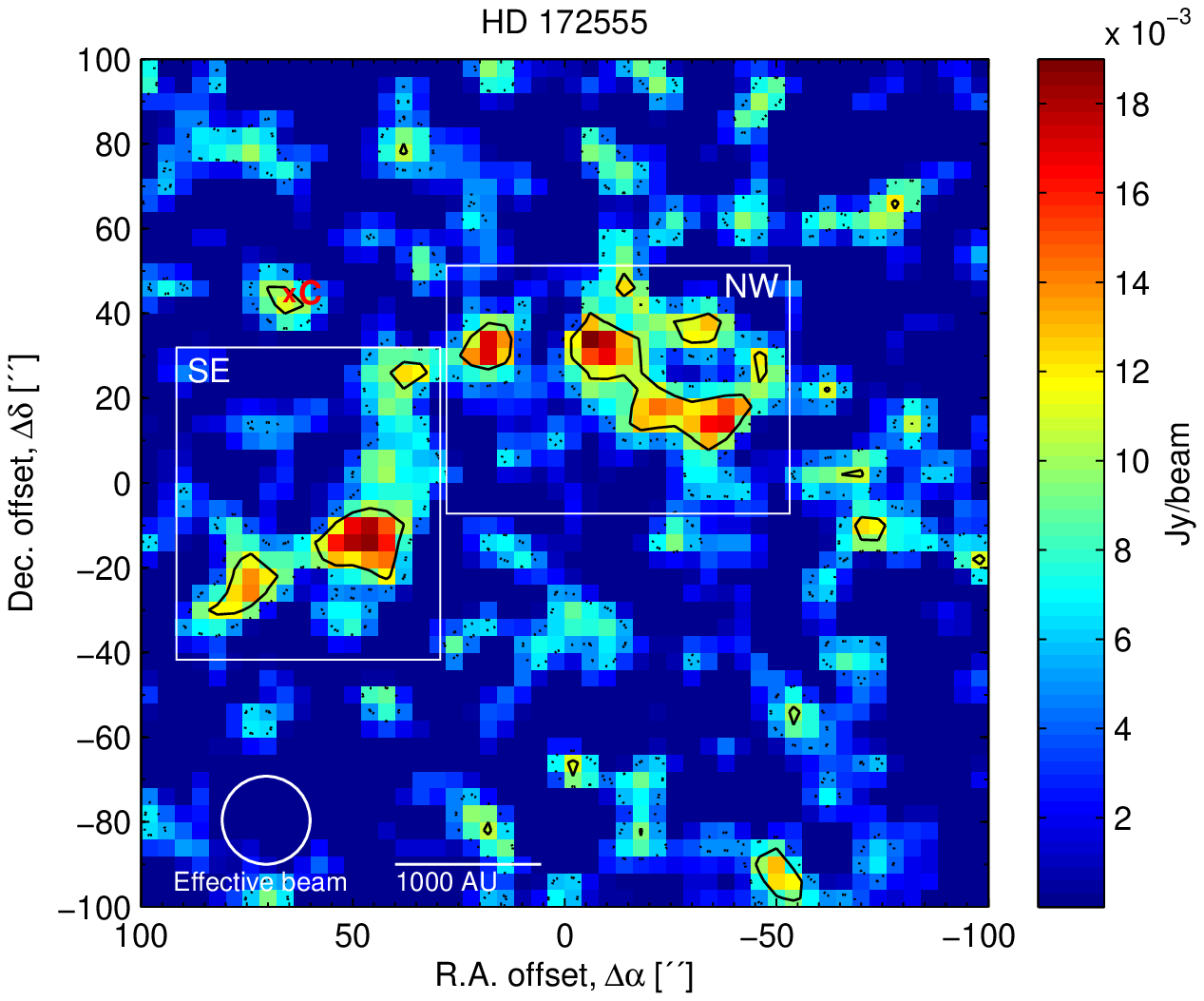}}
     \caption{Maps of $\beta$~Pic, HD\,181327, and HD\,172555 observed at 870\,$\mu$m with LABOCA at the APEX telescope. The maps have an effective resolution of 20{$\farcs$}8, as shown by the circle. In addition, a scale bar corresponding to 1000\,AU at the distance of the star has been inserted. The first solid contour represents 2$\sigma$ flux levels, with the following contours at increments of 1$\sigma$ in (b) and (c), and 2$\sigma$ in (a), while the dotted contour outlines the 1$\sigma$ level. The position of the star, as given in SIMBAD, is at $(\Delta\alpha,\Delta\delta)=(0\arcsec,0\arcsec)$, with additional crosses labelled A and B in subfigure (a) representing the ``blobs'' found by \citet{Holland1998} and \citet{Liseau2003}, respectively. In subfigure (c) the cross labelled C shows the position of the companion star.}
     \label{fig:1all}
\end{figure*}

HD\,181327 is clearly detected and marginally resolved with a fitted FWHM of $31{\arcsec}.9$ (Fig.~\ref{fig:1all}b). The 24.2\,mJy/beam peak flux density is positioned on the star (within the pointing accuracy) with a 3$\sigma$ peak located 22$\arcsec$ away at a PA of 107$\pm$7$\degr$ (with the uncertainty determined as the angle subtended by the pointing accuracy $\pm4\arcsec$ at the peak centre). A 4$\sigma$ peak is also found on the opposite north-west (NW) extension of the disk, thus, the extended submm features are consistent with the  orientation of the inclined dust ring previously imaged in scattered light by HST/NICMOS \citep[PA of 107$\pm$2$\degr$;][]{Schneider2006}. In addition to the emission along the disk, we find a marginally resolved feature orthogonal to the disk, in the southward direction.

\subsection{HD\,172555}
The intriguing 2--3$\sigma$ structure imaged in the vicinity of HD\,172555 could be a very extended dust distribution associated with the star.  Considering that a gravitationally bound companion, the M0 star CD\,-64\,1208, is located at $\mathrm{PA}=56{\degr}$ (almost on a 2$\sigma$ flux-peak, see cross labelled C in Fig.~\ref{fig:1all}c) with a projected separation of 2000\,AU (Zuckerman et al. 2001a), we see that the submm flux density distribution displays a striking symmetry relative to the binary system. The integrated flux density is 30$\pm$10\,mJy for the SE part of the feature, compared to and 40$\pm$10\,mJy for the NW side, calculated within the rectangles enclosing the brighter features in Fig.~\ref{fig:1all}c. The strongest flux density peaks on each side, located at offsets (46$\arcsec$,-14$\arcsec$) and (-6$\arcsec$,34$\arcsec$), both have a brightness of 19\,mJy/beam.

\begin{table*}
\caption{Integrated flux density, root-mean-square noise levels, and derived dust temperature, power law exponent of the opacity law, mass, and fractional dust luminosity for the 870-$\mu$m observations of main-sequence members of the BPMG.}
\label{table:2}
\centering
\begin{minipage}{\textwidth}
\centering
\begin{tabular}{l c c c c c c}
\hline\hline
Target & Integrated Flux Density\footnote{The integrated flux density was calculated by fitting a Gaussian peak for HD\,181327, $\beta$~Pic, and feature II in the $\beta$~Pic map, while a circle of radius $47\arcsec$ was used when including feature I. For HD\,172555 the flux was measured within the boxes in Fig.~\ref{fig:1all}c, as described in the text. Estimated errors come from RMS noise calculation in the specific integration region, together with an absolute calibration error of 10\% \citep{Siringo2009}.} $F$ & RMS Noise, $\sigma$ & $T_{\mathrm{dust}}$\footnote{For detected sources we use the dust temperature derived from the best fit of the spectral energy distribution. For HD\,15115 the best-fit temperature from \citet{Williams2006} is employed, while temperatures from IR excess measured with IRAS \citep{Rhee2007} is used for the other objects.} & $\beta$ & Dust Mass\footnote{Upper $3\sigma$ limit on the dust mass for undetected sources.}, $M_\mathrm{dust}$ & $f_\mathrm{dust}$ \\
 & (mJy) & (mJy/beam) & (K) &  & ($M_{\mathrm{Moon}}$) & ($10^{-4}$) \\
\hline
$\beta$~Pic & 63.6$\pm$6.7 & 2.4 & 89 & 0.67 & 4.8$\pm$0.5 & 17 \\

$\beta$~Pic (incl.~I) & 70.6$\pm$8.2 &  & ... & ... & ... & ... \\

$\beta$~Pic (only II) & 15.8$\pm$2.4 &  & ... & ... & ... & ... \\

HD\,181327 & 51.7$\pm$6.2 & 5.0 & 70 & 0.15 & 34$\pm$4 & 27 \\

HD\,172555 & 30$\pm$10 (SE)\footnote{South-east and north-west feature of HD~172555 calculated separately.} & 5.2 & 320+(10--20) & 0.1+0.1 & 10--60 (SE) & 10\footnote{For a 15\,K disk.}\\

 & 40$\pm$10 (NW) &  & & & 20--70 (NW) & \\

HD\,15115 & ... & 5.1 & 62 & 0.73\footnote{From detection at 850\,$\mu$m by \citet{Williams2006} who found $F_{850}=4.9\pm1.6$\,mJy and derived a dust mass of 3.8\,$M_\mathrm{Moon}$.} & $<2.9$ & 5.8\mpfootnotemark[6]\\

HD\,164249 & ... & 4.0 & 70 & ... & $<2.2$ & 5.9\footnote{From \citet{Rebull2008}.}\\

HD\,181296 & ... & 4.8 & 150 & ... & $<1.3$ & 2.4\mpfootnotemark[7] \\

HD\,191089 & ... & 6.0 & 95 & ... & $<3.2$ & 19\footnote{From \citet{Moor2006}.} \\
\hline
\end{tabular}
\end{minipage}
\end{table*}

\section{Discussion}\label{sec:dis}
\subsection{Warm and Cold Debris Disks in the BPMG}
Together with the previously detected submm emission from $\beta$~Pic \citep{Holland1998} and AU\,Mic \citep{Liu2004}, our observations increase the frequency of detected submm disks in the BPMG to almost 17\% (5 out of 30 stars). IR observations with the Multiband Imaging Photometer for \emph{Spitzer} (MIPS) of this nearby stellar association has suggested a debris disk fraction of more than 37\%, with 7 members showing a 24\,{$\mu$}m excess, at least 11 with a 70\,{$\mu$}m excess, and at least 5 with 160\,{$\mu$}m excess \citep{Rebull2008}. This can be compared with the ${\sim}25$\% disk frequency of nearby stars estimated by \citet{Matthews2007} based on results from IRAS \citep{Backman1993} and \emph{Spitzer} \citep{Rhee2007}. Considering the young age (${\sim}12$\,Myr) of the stars in the BPMG association, the high frequency of disks detected both at IR and submm wavelengths is not surprising. The indication that a large population of cold disks may only be detectable at submm wavelenghts \citep{Lestrade2006,Wyatt2003} also argues for additional observations of members not showing a clear IR excess. The decrease in the fraction of disks detected at wavelengths longward of ${\sim}100$\,$\mu$m is not necessarily an indication of fewer cold disks, but more likely a result of lower detection sensitivities, due to decreased flux densities at the Rayleigh-Jeans tail of the SED combined with lower instrument sensitivities at these wavelengths. Many debris disks may have evolved into systems with a very cold and low density dust distribution, in which an occasional onset of collisional dust avalanches from collisions of planetesimals or comets \citep{Grigorieva2007b}, in structures similar to the Kuiper-Belt or the inner Oort-cloud, can temporarily increase the amount of small particles emitting at submm wavelengths. The stochastic nature of such events could explain the wide scatter in observed dust masses in disks of the same age and with similar spectral types. However, $M_\mathrm{dust}$ carries larger uncertanties than the flux-derived errors indicated in Table \ref{table:2}, due to uncertanties in stellar distance, dust opacity and temperature distribution. A more independent value is given by the fractional dust luminosity, $f_\mathrm{dust}=L_\mathrm{dust}/L_{*}$, which relies only on the flux measurements (see section~\ref{sec:fracdust}).

\subsection{Spectral Energy Distribution and Disk Properties}\label{sec:dis_sed}
\begin{figure*}
     \centering
     \subfigure[]{
     			\label{fig:2a}
          \includegraphics[width=.45\textwidth]{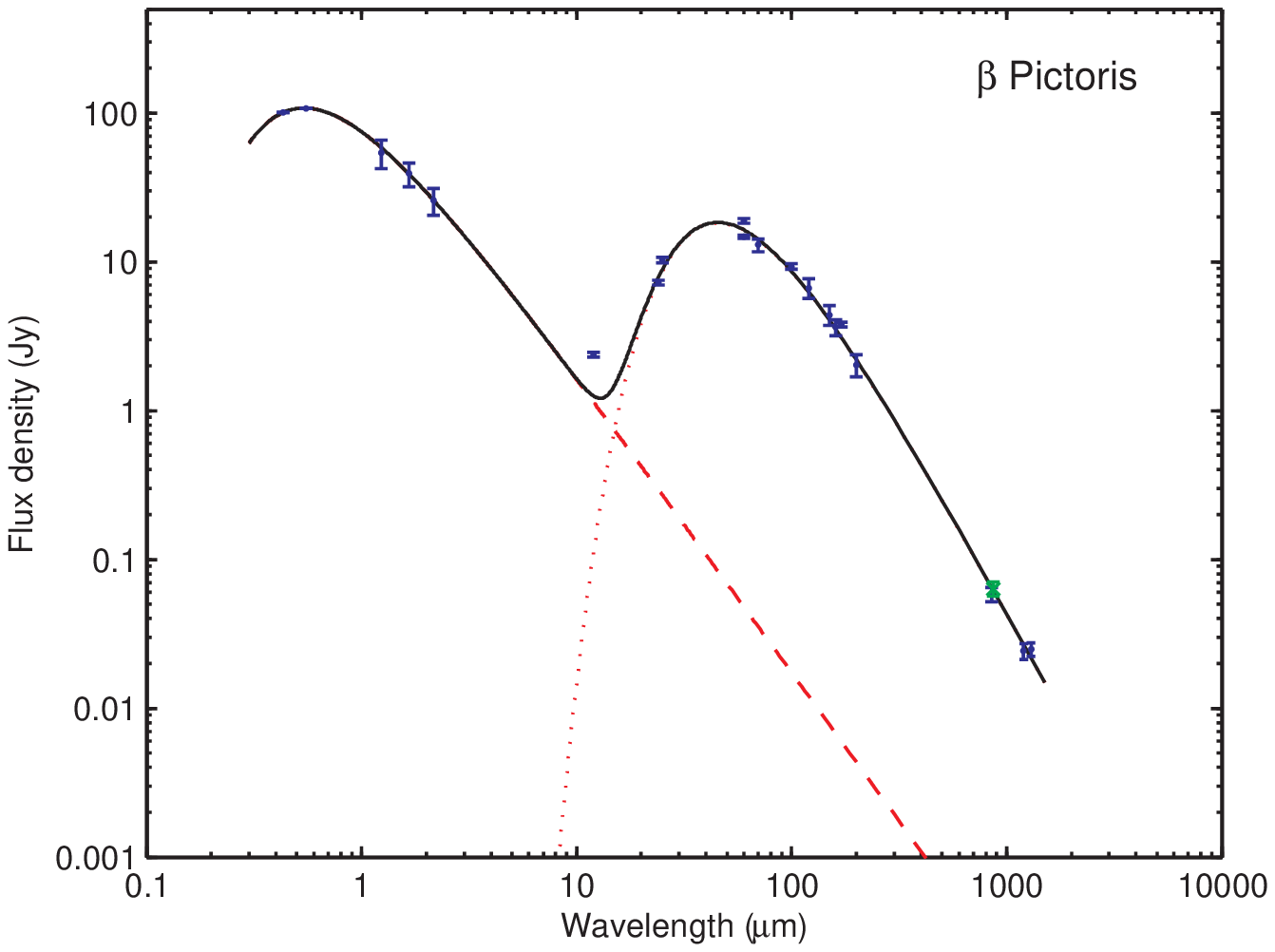}}
     \hspace{1mm}
     \subfigure[]{
          \label{fig:2b}
          \includegraphics[width=.45\textwidth]{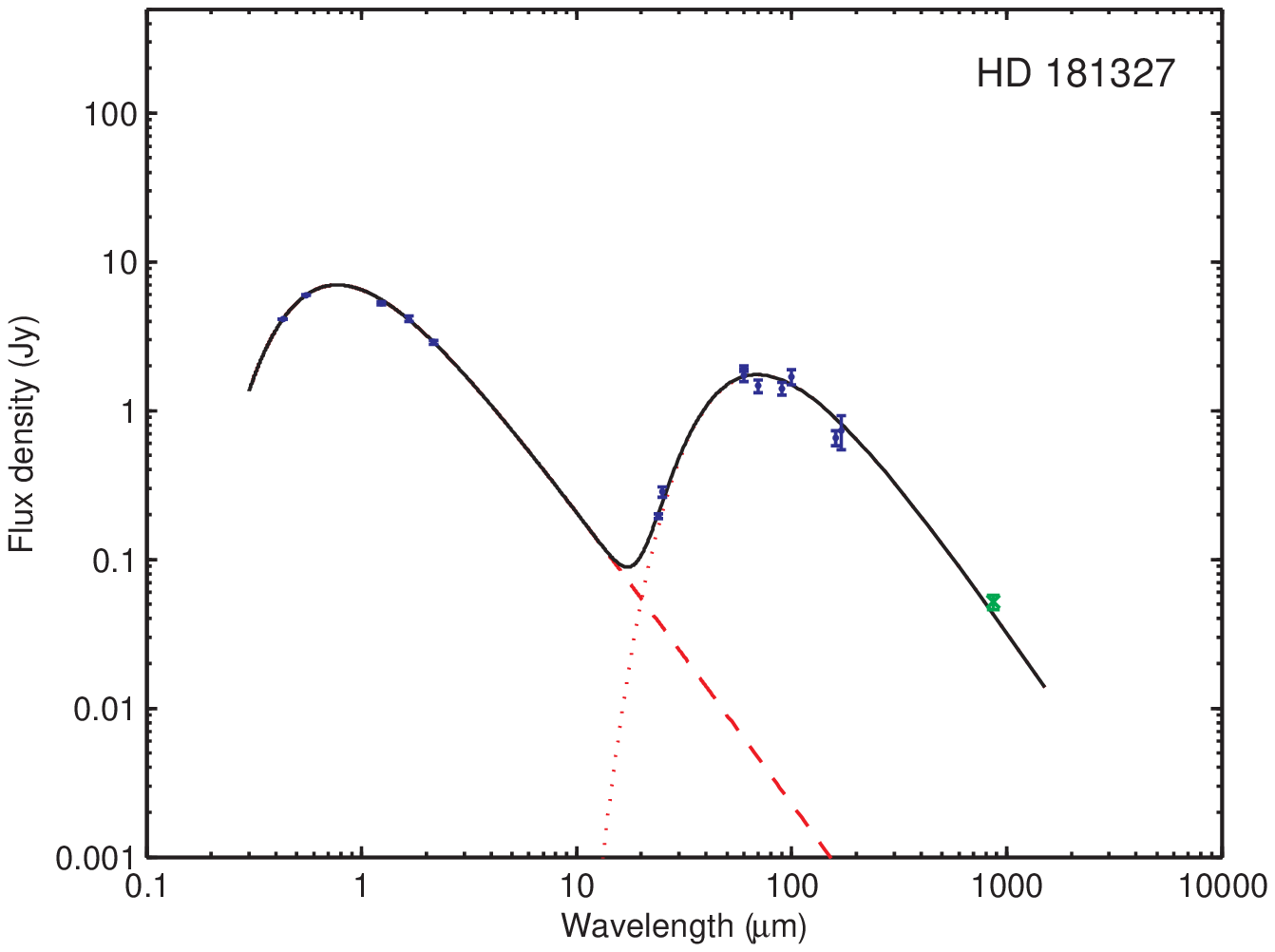}} \\
     \vspace{1mm}
     \subfigure[]{
           \label{fig:2c}
          \includegraphics[width=.45\textwidth]{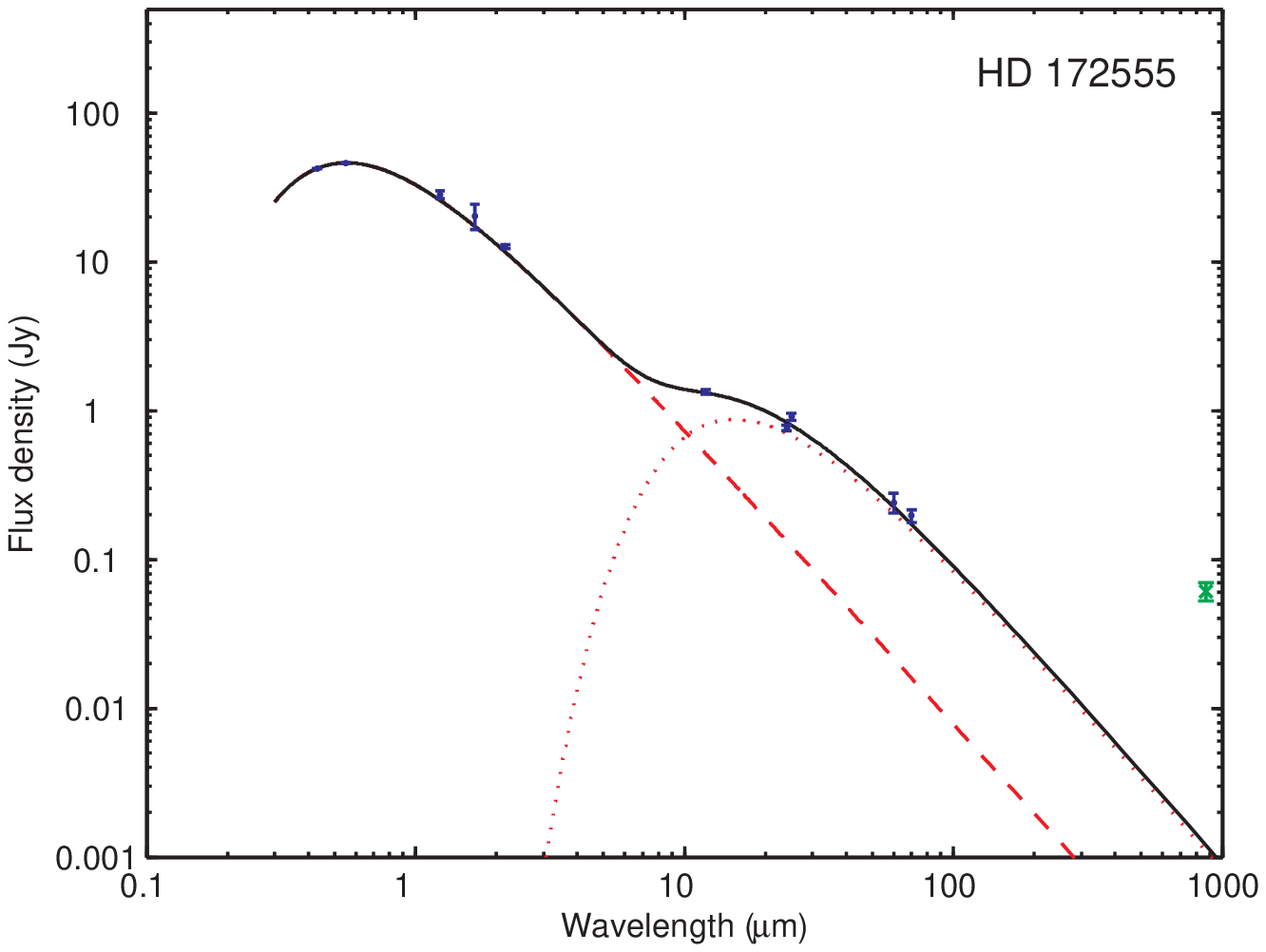}}
     \hspace{1mm}
     \subfigure[]{
          \label{fig:2d}
          \includegraphics[width=.45\textwidth]{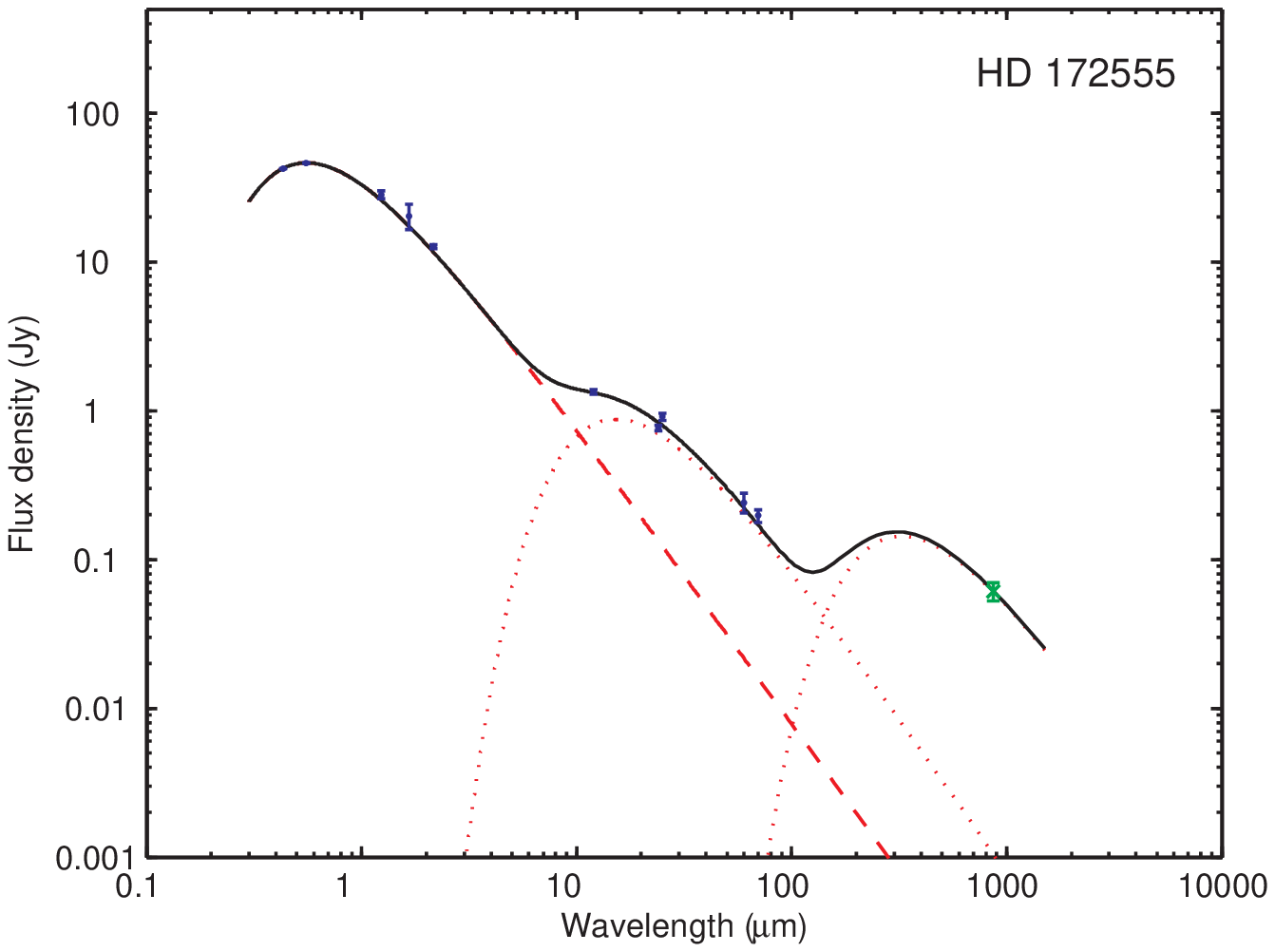}} \\
     \caption{SEDs of the three detected objects in BPMG. Photometry data-points comes from SIMBAD (B and V mag), 2MASS (J,H and K), Spitzer (3.6, 4.5, 5.8, 8.0, 24, 70 and 160\,$\mu$m), IRAS (12, 25, 60 and 100\,$\mu$m), ISO (60, 90, 120, 150, 170 and 200\,$\mu$m), JCMT (850\,$\mu$m), SEST (1200\,$\mu$m), and our new APEX data (870\,$\mu$m) marked with a green cross. A stellar blackbody and a disk modified blackbody has been fit to the data, yielding $T_\mathrm{dust}=89$\,K and $\beta=0.67$ for $\beta$~Pic, $T_\mathrm{dust}=70$\,K and $\beta=0.15$ for HD\,181327, and $T_\mathrm{dust}=322$\,K and $\beta=0.1$ for HD\,172555. Subfigure (d) shows the SED of HD\,172555 fitted with a simplified stellar (blackbody) spectrum, a 320\,K modified blackbody dust spectrum, and a second, colder (15\,K) dust component. The temperature was derived from a simple approximation of blackbody grains at a distance of 1000\,AU from the star. The added dust component must have a temperature below 20\,K in order to match the 870\,$\mu$m data without being inconsistent with far-IR data. See section \ref{sec:dis_sed} for discussion.}
     \label{fig:2all}
\end{figure*}

The measured submm flux densities of $\beta$~Pic, HD\,181327, and HD\,172555 add an important data point to earlier mid- and far-IR photometry, enabling a better determination of the SED of the disk, and thereby its temperature and radial extent. The thermal radiation emitted from a population of dust particles depends on their temperature distribution and opacity. Assuming that the opacity index varies with the frequency as a power law, i.e.~$\kappa_{\nu}=\kappa_{0}(\nu/\nu_{0})^{\beta}$, and that the dust grains dominating the flux in the IR and submm part of the SED all have the same temperature $T_{\mathrm{dust}}$, then the flux density spectral distribution can be approximated with a modified blackbody function of that temperature, as
\begin{equation}
F_{\mathrm{disk}}(\nu)=\frac{2h{\nu}^{3}}{(\mathrm{e}^{\frac{h\nu}{kT_{\mathrm{dust}}}}-1)c^{2}}\left(\frac{\nu}{\nu_{0}}\right)^{\beta}{\kappa}_{0}\Omega=\frac{{\nu}^{3+\beta}}{(\mathrm{e}^{\frac{h\nu}{kT_{\mathrm{dust}}}}-1)}C.
\end{equation}
Naturally, the dust grains dominating the submm emission would be expected to be colder (located at Kuiper-belt distances) than the main IR emitting grains (at asteroid-belt distances), but the assumption of equal dust temperature is a generally adopted simplification due to limited spatial information. The power law exponent, $\beta$, of the opacity law, basically depends on material properties and the size distribution of the dust grains. An exponent ${\beta} \approx 0$ would point to grains radiating almost as blackbodies, while largely unprocessed interstellar dust grains have been shown to have ${\beta} \approx 2$ \citep{Hildebrand1983}. A $\beta$ of 0.5--1 has been found in protoplanetary (T~Tauri) disks, indicating amorphous or fractal dust grains, or perhaps just grains significantly larger than those in the interstellar medium \citep{Mannings1994}.
We made $\chi^{2}$-minimisation fits to available photometry data (weighted with the accuracy of the flux density measurement) of our three detected objects with the total SED function, $F_{\mathrm{SED}}(\nu)=F_{\mathrm{star}}(\nu)C_{\mathrm{star}}+F_{\mathrm{disk}}(\nu)(C_{\mathrm{disk}}/C)$, where the stellar photosphere $F_{\mathrm{star}}$ is simply approximated by a blackbody function, and $C_{\mathrm{star}}$ and $C_{\mathrm{disk}}$ are scale factors depending on distance and radiating surface area. The best fit for each object is displayed in Fig.~\ref{fig:2all}. In Table~\ref{table:3} (available as online material) we have compiled all current mid-IR, far-IR, and submm photometry data of the 12 BPMG members observed in the submm so far.
At submm wavelengths the disk is thought to be optically thin, with the flux coming predominantly from larger grains than the ones emitting in the IR. An estimate of the dust mass can then be found from
\begin{equation}
M_{\mathrm{dust}}=\frac{F_{\nu}D^2}{{\kappa_{\nu}}B_{\nu}(T_{\mathrm{dust}})},
\end{equation}
where $F_{\nu}$ is the integrated 870\,$\mu$m flux density, $D$ is the distance to the star, and $B_{\nu}$ can be approximated with the Rayleigh-Jeans expression, $B_{\nu}={2{\nu^2}kT}/{c^2}$, at long wavelengths. We adopt the value $\kappa=2\,\mathrm{cm^{2}g^{-1}}$ for grain sizes of about $1\,\mathrm{mm}$ at ${\lambda}{\sim}1\,\mathrm{mm}$ (see discussion in \citet{Liseau2008} and \citet{Miyake1993}). Dust temperatures and masses derived from the fits are presented in Table~\ref{table:2}, where values for undetected sources are calculated from dust temperatures measured with IRAS \citep{Rhee2007}.

By assuming that the grains are in thermal equilibrium with their environment, i.e., that the absorbed stellar radiation energy is emitted at the same rate, we can calculate their characteristic radial distance from the star as
\begin{equation}
r=\left(\frac{R}{2}\right)\left(\frac{T_{\mathrm{e}}}{T_{\mathrm{dust}}}\right)^{2}\sqrt{\frac{(1-A)}{\epsilon}}.
\label{eqn:3}
\end{equation}
The reflectance, $A$, and emittance, $\epsilon$, of debris disk grains are currently uncertain, but assuming amorphous grains without ice mantles, we can simplify by approximating them as blackbodies with a temperature corresponding to a fitted Planck-function. This can be justified considering (1) the porous grains required to explain e.g.~polarization maps of AU~Mic \citep{Graham2007}, and outcome from laboratory experiments of dust particle interactions \citep{Wurm1998}; and (2) results from e.g.~the $\beta$~Pic model by \citet{Grigorieva2007} showing that any $\sim$1\,mm sized grains at radial distances out to 150\,AU would, due to photosputtering, be devoid of ice even though located beyond the snow-line.

\subsubsection{$\beta$~Pictoris}\label{sec:dis_pic}
The edge-on debris disk surrounding the nearby A6V star $\beta$~Pic has been intensively studied, with images in optical and near-IR light revealing a disk with $>$1300\,AU radius, and morphological asymmetries indicating dynamical interaction with planets \citep{Kalas1995,Mouillet1997,Heap2000}. Further indications of planets, from inner disk gaps observed in the mid-IR \citep{Lagage1994} was recently followed by near-IR imaging of a possible giant planet \citep{Lagrange2009}. Spectroscopic investigations have found silicate emission features \citep{Chen2007} and variable redshifted metallic lines suggesting falling evaporating cometary bodies \citep{Beust1994}, which could continuously replenish the dust disk and explain many characteristics of the near-IR to millimetre emission \citep{Li1998}. \citet{Holland1998} made the first submm detection at 850\,$\mu$m with SCUBA on the JCMT, in which also the SW ``blob'' was imaged. Our observed submm disk of $\beta$~Pic does not appear as clearly elongated as in the image by \citet{Holland1998}, but the integrated flux density is similar 63.6\,mJy compared to 58.3\,mJy). It is possible that the smaller 14$\arcsec$ FWHM beam of SCUBA at 850\,$\mu$m allowed the disk to be marginally resolved, while the 18$\arcsec$ FWHM beam of LABOCA did not. Furthermore, the (suspected) slightly out-of-focus beam of the present observation would have affected the effective image resolution.

The SW flux peak of \citet{Holland1998} (labelled A in Fig.~\ref{fig:1all}a) is confirmed with a 5$\sigma$ significance peak, centered at the same (within the pointing accuracy of the telescope) position with a PA of $36\pm5\degr$ compared to $37\pm6\degr$, just off the main optical disk \citep[$\mathrm{PA}=30.7\pm0.7\degr$,][]{Kalas1995}, some 600--700\,AU from the star. Modelling by \citet{Dent2000} has shown that if this emission component really comes from dust belonging to the system and is being heated by the star, it must have a mass comparable to the rest of the dust disk and would in that case also be visible in scattered optical light, which it is not \citep[see, e.g.,][]{Golimowski2006}. On the other hand, if the grains are fluffy, as in the comet dust model by \citet{Li1998}, this dust would not have to be visible in scattered light \citep[compare with the dust ring of $\epsilon$~Eri, which is not seen with either HST/STIS or near-IR interferometry with CHARA;][]{Proffitt2004,diFolco2007}. In addition to the discussed flux density peak, another even stronger emission peak is found at the same PA ($36\pm5\degr$) but at a distance of 2000--2500\,AU from the star. The first interpretation of these bright features would be that they are background submm galaxies, but the coincidental alignment with features (like warps) in the observed optical and IR disk is puzzling indeed. As discussed in \citet{Holland1998} the emission could come from primordial disks around younger, low-mass companions to the star. However, in order to have avoided detection in the optical and IR such companions must have a mass closer to gas-giant planets. The location of the feature observed with SIMBA at 1200\,$\mu$m \citep{Liseau2003} does not correspond to any flux exceeding 2$\sigma$ in our 870\,$\mu$m map, however, faint emission can be seen extending southward like in the SIMBA and SCUBA images, reaching a 3$\sigma$ peak on the other side of the disk midplane. This latter flux peak is located at a distance comparable to where the most remote dust scattered light has been observed \citep[$\sim$1400\,AU,][]{Larwood2001} on the SW side of the disk. 
 
For the source centered flux the best fit (see Fig.~\ref{fig:2all}a) yields a dust temperature of 89\,K (giving a dust mass of 4.8\,$M_{\mathrm{Moon}}$) and power law exponent of the opacity law $\beta=0.67$. This temperature is considerably lower than the 130\,K derived in a similar simple model by \citet{Rebull2008} but then only considering ${\lambda}{\leq}160\,\mu$m. In addition to our own new submm data, we included the 1.2\,mm detection by \citet{Liseau2003}, who also arrived at an equally shallow opacity index. The dust in the disk of $\beta$~Pic would be expected to have a low $\beta$ as a sign of big grains. From (\ref{eqn:3}) we get the dust grains characteristic distance from the star, which becomes $r=40$\,AU. This falls in the region where dust rings have previously been imaged at 17.9\,$\mu$m \citep{Wahhaj2003}, and where the Kuiper-belt is located in our Solar system \citep[e.g.][]{Weissman1995}. However, this is also a region which, according to most models, is dust depleted \citep[with a peak at $\sim$100--120\,AU and a relative inner hole inside 50\,AU;][]{Augereau2001}. This is an interesting paradox, which goes beyond the scope of the present paper, but would have to be explained within the frame of a global coherent model of the $\beta$~Pic system.

\subsubsection{HD\,181327}
The F5/F6V star HD\,181327 has, as mentioned earlier, been found by HST/NICMOS coronagraphic imaging to have an inclined circumstellar ring of dust. Located at a radial distance of 86\,AU from the star and being 36\,AU wide \citep{Schneider2006} this feature is currently not possible to spatially resolve with single-dish telescopes at submm wavelengths since an angular resolution $<2\arcsec$ would be required. However, Schneider et al.'s follow-up observations with HST/ACS revealed a diffuse scattering halo out to $\sim$500\,AU, indicating that more extended dust surrounds the star. Although our submm image at best marginally resolves the source, we do notice an elongation in a direction consistent with the PA of the scattering ring, as well as southward, almost orthogonal to the disk.  
In Fig.~\ref{fig:2all}b we see that the 870\,$\mu$m point lies above the best-fit SED, which has $T_{\mathrm{dust}}=70$\,K and $\beta=0.15$. This suggests that a different and colder population of grains is dominating the excess emission in the submm part of the spectrum than that responsible for the previously detected IR excess in IRAS observations \citep{Mannings1998} and Spitzer IRS spectra \citep{Chen2006}. The existence of an icy Kuiper Belt around HD\,181327 was recently predicted by \citet{Chen2008} based on Gemini South $Q_a$-band 18.3-{$\mu$}m imaging and \emph{Spitzer} MIPS SED-mode observations. Using the higher temperature derived from the fit of the modified blackbody function we get a total dust mass of 34\,$M_{\mathrm{Moon}}$, which is remarkably high compared to the few lunar masses found in most debris disks.

\subsubsection{HD\,172555}
We do not detect any significant submm flux at the position of the star, however, the observed features in the region nearby and between HD\,172555 and its companion star does appear source related. The orbital inclination of the A7/M0 binary system is unknown, but in order for the observed structure to come from dust in bound orbits, it must be located within $\sim$1/3 of the binary distance from the primary \citep[see e.g.,][]{Cuntz2007}, thus the viewing angle must be fairly high. It is difficult to draw conclusions on the geometry of the dust distribution, but the similarity in the total integrated flux within the NW and SE features, respectively, and also between the flux densities of their brightest features in Fig.~\ref{fig:1all}c could point to dust being produced by collisions of cometary bodies at the inner rim of a structure comparable to the proposed dense inner doughnut-shaped part of the Oort cloud \citep{Hills1981,Levison2001}. Although the observed flux distribution could also originate from e.g., two interacting background galaxies, the probability of such a chance alignment must be considered small. The initial IRAS detection of this IR excess source revealed an anomalously high dust temperature of 320\,K \citep{Oudmaijer1992,Mannings1998}, and \emph{Spitzer} IRS spectroscopy has shown warm silicate dust features \citep{Chen2006} and signs of a recent hypervelocity collision between two large bodies \citep{Lisse2009}. Modelling by \citet{Wyatt2007} suggests that the dust luminosity is too high to be explained by steady state collisional processing of planetesimal belts, and that HD\,172555 could be undergoing a transient event. Thus, this system may be experiencing a period of heavy bombardment, with frequent collisions of planetesimal bodies which could explain the high mass of localized dust at 1000\,AU required to explain our observations, and the absence of large, cold dust grains at Kuiper-belt distances.
The main emitting regions are, as mentioned, located at a projected distance of $\sim$1000\,AU from HD\,172555. Consequently, again assuming thermal equilibrium, we can use (\ref{eqn:3}) to find the expected temperature of the dust grains. With the blackbody approximation we arrive at $T_{\mathrm{dust}}=15\,\mathrm{K}$. In addition, comparing the position of the 870\,$\mu$m flux point to the best-fit curve in Fig.~\ref{fig:2all}c, which has $T_{\mathrm{dust}}=322$\,K and $\beta=0.10$, it can be clearly seen that the detected dust must be very cold and that the bulk of the submm emission does not come from the same population of grains observed in the IR. A second (blackbody) dust component with $T_{\mathrm{dust}}=15$\,K can also be nicely fitted with the SED data (see Fig.~\ref{fig:2d}). Although we have basically only one data point sampling this colder dust, we note that anything above 20\,K would be difficult to match to the far-IR data, while a significant dust population with a temperature below 10\,K would be surprising at these distances. This range of temperatures has been used to estimate the lower and upper limit on the dust mass presented in Table \ref{table:2}. More detailed modelling of the extended flux would allow us to determine the geometry and orientation of the emitting dust structure.

For both HD\,172555 and HD\,181327 additional photometry at slightly shorter (e.g.~450\,$\mu$m with SABOCA at APEX) and longer (millimetre with ALMA) wavelengths are important to constrain the temperature and nature of the observed colder dust component.

\subsection{Fractional Dust Luminosity and Temporal Evolution of Debris Disks}\label{sec:fracdust}
\begin{figure}
    \includegraphics[width=0.5\textwidth]{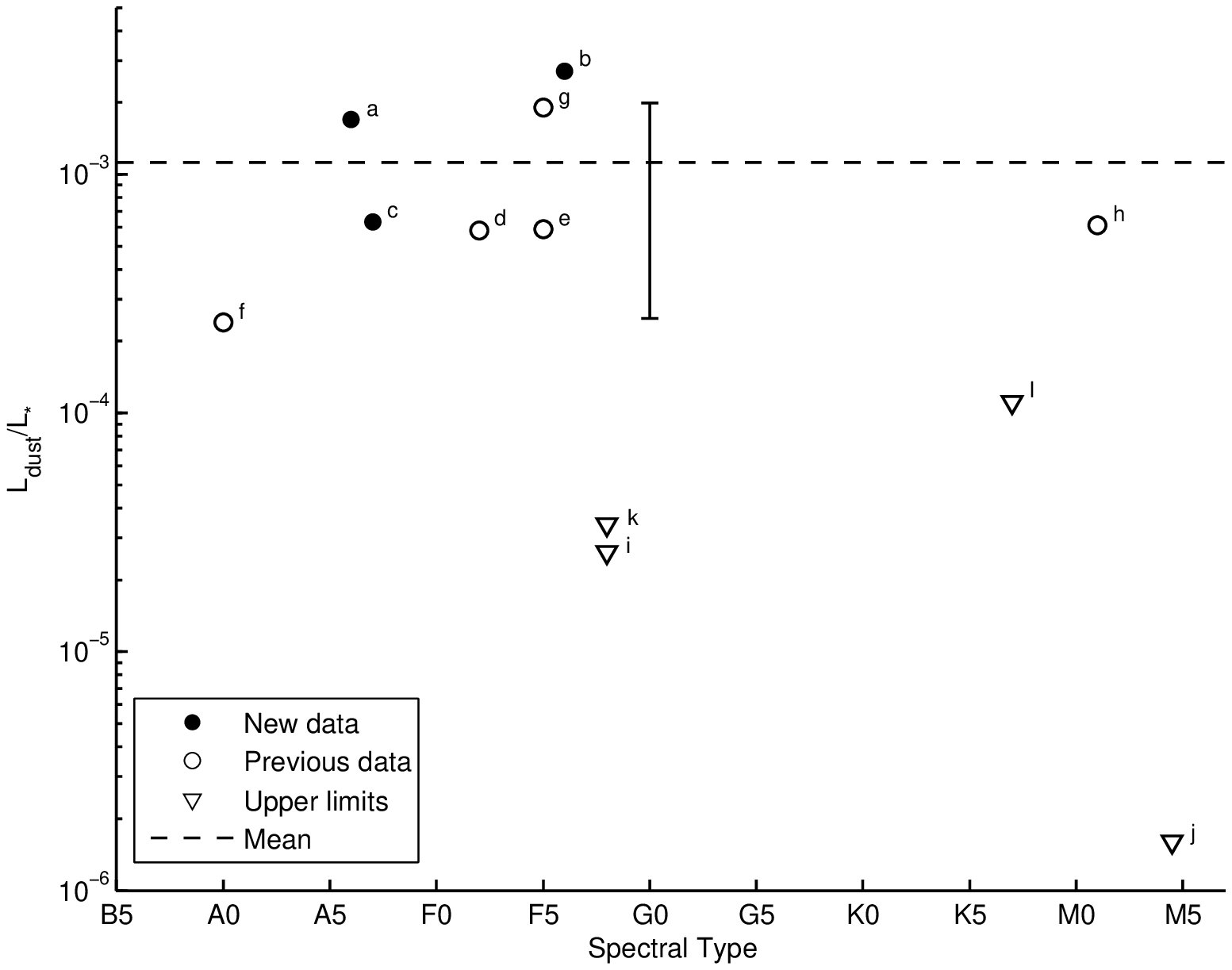}
    \vspace{0cm}
    \caption{Plot of fractional dust luminosities of debris disks from fitting SEDs to available flux data of stars in the $\beta$~Pic Moving Group, versus stellar spectral type. It shows our detections (filled circles): (a) $\beta$~Pic, (b) HD\,181327, (c) HD\,172555, together with previous results (open circles): (d) HD\,15115 \citep{Williams2006}, (e) HD\,164249 \citep{Rebull2008}, (f) HD\,181296 \citep{Rebull2008}, (g) HD\,191089 \citep{Moor2006}, (h) HD\,197489 (AU~Mic) \citep{Liu2004}, and upper limits (open triangles): (i) HD\,35850 (AF~Lep), (j) HD\,196982 (AT~Mic), (k) HD\,199143, and (l) HD\,358623 (AZ~Cap). The upper limits were found from SED fitting of upper ($3\sigma$) flux limits (see Table~\ref{table:3} for data and references). For HD\,172555 we plotted a fractional dust luminosity calculated for a second dust component with $T_\mathrm{dust}=15$\,K. Mean $f_\mathrm{dust}$ (dashed line) and standard deviation (error bar) are also shown.}
    \label{fig:3}
\end{figure}

Using our fitted SEDs and previous measurements by \citet{Williams2006}, \citet{Rebull2008}, \citet{Moor2006}, and \citet{Liu2004}, we can make a plot of $f_\mathrm{dust}$ versus spectral type for all currently known debris disks in the BPMG (see Fig.~\ref{fig:3}). The mean $f_\mathrm{dust}$ is $11\cdot10^{-4}$, with a large standard deviation $\sigma=9\cdot10^{-4}$. HD\,35850, HD\,196982, HD\,199143, and HD\,358623 all show exceptionally low upper limits, which could be due to their special nature; the first being variable, the second and third being binary, and the fourth being both. Dynamical interactions in binary systems would lead to smaller stability zones for debris disks and possibly a faster decline in fractional dust luminosity \citep{Trilling2007,Holman1999}.

A rough comparison of the ${\sim}12\,$Myr BPMG with the upper limit, $f_\mathrm{dust}<2\cdot10^{-4}$, derived for the ${\sim}100\,$Myr Pleiades cluster \citep{Greaves2009} suggests an $f_\mathrm{dust} \propto M_\mathrm{dust} \propto t^{-\alpha}$ decline of the emitting dust, with ${\alpha}>0.8$. Early results from ISO observations of clusters of various ages gave $\alpha=2$ \citep{Spangler2001}, while more recent Spitzer and JCMT data have hinted at a more modest decline with $\alpha=0.6$ \citep{Su2006} and $\alpha=0.5-1$ \citep{Liu2004}, respectively. Models of collisional dust evolution in debris disks by \citet{Wyatt2007} are consistent with these results, showing an $f_\mathrm{dust}$ of 5--$10{\cdot}10^{-4}$ at $t \approx 10$\,Myr, with a falloff as $t^{-0.4}$. The relatively large spread of $f_\mathrm{dust}$ also for stars of similar spectral type (and mass) found in BPMG (A and F stars) could perhaps be explained by the models of \citet{Kenyon2008}, which show that 10-\,Myr old A--F type stars with a 30--150\,AU disk could be just at the steeply rising slope of the fractional dust luminosity created by the onset of collisional cascades. The rise from $f_\mathrm{dust} \approx 0.1{\cdot}10^{-4}$ to $100{\cdot}10^{-4}$ would come earlier around the more massive stars and the stars with more massive disks, perhaps as early as 0.15\,Myr for a $3\,M_{\sun}$ mass star with a 3\,MMSN disk, and as late as 10\,Myr for a $1\,M_{\sun}$ mass star with a 1/3\,MMSN disk (see Figs.~14 and 17 in \citet{Kenyon2008}).

\addtocounter{table}{1}

\section{Conclusions}
We summarize the most important findings of our 870\,$\mu$m observations of 7 main-sequence IR-excess stars in the BPMG as follows:
   \begin{itemize}
      \item[\textbullet] Out of the 7 stars observed, we made three detections. Two of the detected objects have previously never been detected at submm wavelengths. Our observations increase the frequency of detected submm disks in the BPMG to almost 17\% (5 out of 30 stars).
      \item[\textbullet] $\beta$~Pic showed a strong flux density peak centered on the star, with two additional weaker flux peaks, both at a PA of $\sim36\degr$ and radial distances of 600--700\,AU and 2000--2500\,AU, respectively. The former of these has a position consistent with the ``blob'' imaged by \citet{Holland1998} at 850\,$\mu$m. Both may be background submm galaxies, but surprisingly show a close alignment to the observed optical and IR disk plane. Simple fitting of a stellar blackbody and a disk modified blackbody to the SED data suggests a dust temperature of 89\,K and a low opacity law exponent $\beta \approx 0.7$. We calculate a minimum dust mass using the theoretically derived maximum opacity index for 1\,mm sized icy grains and arrive at 4.8\,$M_{\mathrm{Moon}}$.
      \item[\textbullet] HD\,181327 is clearly detected but at most marginally resolved with an elongation consistent with the previously imaged inclined circumstellar dust ring. SED fitting gives $T_{\mathrm{dust}}=70$\,K and $\beta=0.15$, but with the 870\,$\mu$m flux ending up above the fit, implying the existence of a population of colder dust grains. The calculated minimum dust mass $M_\mathrm{dust}=34\,M_{\mathrm{Moon}}$ is remarkably high. 
      \item[\textbullet] HD\,172555 does not show any significant submm emission at the position of the star, however, in the region between the star and its lower mass companion a very extended flux density distribution is seen. Assuming that the emission comes from dust associated with HD\,172555 we derive the thermal equilibrium temperature of dust at the projected distances ($\sim$1000\,AU), which becomes 15\,K. The integrated flux in the observed features lies far above what would be expected for a single modified blackbody disk fit to IR photometry data in the SED. Adding a second disk component with a temperature of 10--20\,K fits nicely with the observed SED, and agrees with the temperature estimated from dust grain distances to the star. The corresponding dust mass would be 10--60\,$M_{\mathrm{Moon}}$ and 20--70\,$M_{\mathrm{Moon}}$ for the NW and SE feature, respectively.
      \item[\textbullet] We derive the fractional dust luminosity of detected sources, which is then plotted versus stellar spectral type. The mean fractional dust luminosity, $\bar{f}_\mathrm{dust}=11{\cdot}10^{-4}$ measured for debris disk in the BPMG, agrees with the expected stage of collisional dust evolution for such ${\sim}12\,$Myr systems predicted from previous observations \citep[e.g.][]{Su2006,Liu2004,Spangler2001} and models \citep{Kenyon2008,Wyatt2007}. The large scatter in $f_\mathrm{dust}$ among co-eval stars could be a sign of stochastic collisional dust production or a consequence of BMPG's evolutionary stage, perhaps just at the onset of collisional dust cascades at $t \approx 10\,$Myr. Comparison with data at 100\,Myr \citep{Greaves2009} gives $f_\mathrm{dust} \propto t^{-\alpha}$ with ${\alpha}>0.8$.
   \end{itemize}

\begin{acknowledgements}
      We are grateful for instructive discussions with Philippe Th\'{e}bault, and would also like to thank Andreas A.~Lundgren for helpful advice regarding the data reduction, and Attila Kovacs for great technical support of the MiniCRUSH software. This research was contributed to by financial support from Stockholm Astrobiology Graduate School, and the International Space Science Institute (ISSI) in Bern, Switzerland (``Exozodiacal Dust Disks and Darwin'' working group, http://www.issibern.ch/teams/exodust/). AB was funded by the \textit{Swedish National Space Board} (contract 84/08:1). Lastly, we thankfully acknowledge the comments by the anonymous referee, which helped improve the paper.
\end{acknowledgements}

\bibliographystyle{aa}
\bibliography{myref} 

\Online
\begin{appendix}
\longtab{3}{
\begin{longtable}{l c c c c c}
 \caption{\label{table:3} Mid-IR, far-IR, and submm photometry of all BPMG members that have been observed at submm wavelengths. A $3\sigma$ upper limit on the flux density is given for non-detections.}\\
\hline\hline
Star & Other Name & Instrument & $\lambda$ & $F_{\nu}$ & Reference\\
 &  &  & ($\mu$m) & (Jy) &  \\
 \hline
\endfirsthead

\multicolumn{6}{c}{{\tablename} \thetable{} -- Continued} \\
\hline\hline
Star & Other Name & Instrument & $\lambda$ & $F_{\nu}$ & Reference\\
 &  &  & ($\mu$m) & (Jy) &  \\
 \hline
\endhead

HD\,15115 &  & MIPS & 24 & $0.054\pm0.006$ & 1\\

 &  & IRAS & 60 & $0.473\pm0.052$ & 1\\

 &  & ISOPHOT & 60 & $0.415\pm0.020$ & 1\\

 &  & ISOPHOT & 90 & $0.427\pm0.030$ & 1\\

 &  & SCUBA & 850 & $0.0049\pm0.0016$ & 2\\
 
 &  & LABOCA & 870 & $<0.0153$ & 3\\

HD\,35850 & AF~Lep & MIPS & 24 & $0.0790\pm0.00316$ & 4\\ 

 &  & MIPS & 70 & $0.0447\pm0.00447$ & 4\\

 &  & MIPS & 160 & $<0.077$ & 4\\

 &  & IRAS & 12 & $0.4900\pm0.0588$ & 5, 6\\

 &  & IRAS & 25 & $0.080\pm0.008$ & 5, 6\\
 
 &  & IRAS & 60 & $<0.180$ & 5, 6\\

 &  & IRAS & 100 & $<1.930$ & 5, 6\\

 &  & ISOPHOT & 60 & $0.049\pm0.014$ & 5, 6\\

 &  & ISOPHOT & 90 & $0.050\pm0.012$ & 5, 6\\

 &  & SCUBA & 850 & $<0.0045$ & 7\\
 
HD\,39060 & $\beta$~Pic & MIPS & 24 & $7.276\pm0.29104$ & 4\\

 &  & MIPS & 70 & $12.990\pm1.299$ & 4\\

 &  & MIPS & 160 & $3.646\pm0.43752$ & 4\\
 
 &  & IRAS & 12 & $2.38\pm0.10$ & 1\\

 &  & IRAS & 25 & $10.29\pm0.41$ & 1\\
 
 &  & IRAS & 60 & $18.92\pm0.76$ & 1\\

 &  & IRAS & 100 & $9.32\pm0.37$ & 1\\
 
 &  & ISOPHOT & 60 & $14.7\pm0.346$ & 8\\
 
 &  & ISOPHOT & 120 & $6.680\pm1.010$ & 1\\
 
 &  & ISOPHOT & 150 & $4.390\pm0.670$ & 1\\
 
 &  & ISOPHOT & 170 & $3.807\pm0.143$ & 1\\
 
 &  & ISOPHOT & 200 & $2.030\pm0.340$ & 1\\
 
 &  & SCUBA & 850 & $0.0583\pm0.0065$ & 9\\
 
 &  & LABOCA & 870 & $0.0636\pm0.0067$ & 3\\
 
 &  & SIMBA & 1200 & $0.0243\pm0.003$ & 10\\
 
 &  & SEST & 1300 & $0.0249\pm0.0026$ & 16\\
 
HD\,164249 &  & MIPS & 24 & $0.076\pm0.0304$ & 4\\ 

 &  & MIPS & 70 & $0.624\pm0.0624$ & 4\\

 &  & MIPS & 160 & $0.104\pm0.01248$ & 4\\
 
 &  & IRAS & 12 & $0.280\pm0.028$ & 11\\

 &  & IRAS & 25 & $<0.158\pm$ & 11\\
 
 &  & IRAS & 60 & $0.669\pm0.094$ & 1\\

 &  & IRAS & 100 & $<2.63$ & 11\\
 
 &  & ISOPHOT & 60 & $0.761\pm0.0380$ & 1\\
 
 &  & ISOPHOT & 90 & $0.560\pm0.0390$ & 1\\
 
 &  & LABOCA & 870 & $<0.012$ & 3\\

HD\,172555 &  & MIPS & 24 & $0.766\pm0.03064$ & 4\\ 

 &  & MIPS & 70 & $0.197\pm0.0197$ & 4\\
 
 &  & IRAS & 12 & $1.34\pm0.05$ & 1\\

 &  & IRAS & 25 & $0.91\pm0.05$ & 1\\
 
 &  & IRAS & 60 & $0.241\pm0.036$ & 1\\

 &  & LABOCA & 870 & $0.07\pm0.01$ & 3\\

HD\,181296 & $\eta$~Tel & MIPS & 24 & $0.382\pm0.01528$ & 4\\

 &  & MIPS & 70 & $0.409\pm0.0409$ & 4\\

 &  & MIPS & 160 & $0.068\pm0.00816$ & 4\\
 
 &  & IRAS & 12 & $0.351\pm0.018$ & 1\\

 &  & IRAS & 25 & $0.496\pm0.025$ & 1\\
 
 &  & IRAS & 60 & $0.449\pm0.040$ & 1\\
 
 &  & ISOPHOT & 60 & $0.433\pm0.022$ & 1\\
 
 &  & ISOPHOT & 90 & $0.286\pm0.021$ & 1\\
 
 &  & LABOCA & 870 & $<0.0144$ & 3\\

HD\,181327 &  & MIPS & 24 & $0.195\pm0.0078$ & 4\\

 &  & MIPS & 70 & $1.468\pm0.1468$ & 4\\

 &  & MIPS & 160 & $0.658\pm0.07896$ & 4\\
 
 &  & IRAS & 12 & $0.164\pm0.016$ & 11\\

 &  & IRAS & 25 & $0.286\pm0.023$ & 1\\
 
 &  & IRAS & 60 & $1.93\pm0.08$ & 1\\

 &  & IRAS & 100 & $1.69\pm0.20$ & 1\\
 
 &  & ISOPHOT & 60 & $1.730\pm0.170$ & 1\\
 
 &  & ISOPHOT & 90 & $1.410\pm0.140$ & 1\\

 &  & ISOPHOT & 170 & $0.736\pm0.192$ & 1\\
 
 &  & LABOCA & 870 & $0.0517\pm0.0062$ & 3\\

HD\,191089 &  & MIPS & 24 & $0.1856\pm0.0076$ & 12\\

 &  & MIPS & 70 & $0.5443\pm0.0401$ & 12\\

 &  & MIPS & 160 & $0.2046\pm0.0446$ & 12\\
 
 &  & IRAS & 25 & $0.387\pm0.058$ & 1\\
 
 &  & IRAS & 60 & $0.729\pm0.058$ & 1\\
 
 &  & ISOPHOT & 60 & $0.781\pm0.081$ & 1\\
 
 &  & ISOPHOT & 90 & $0.370\pm0.038$ & 1\\
 
 &  & LABOCA & 870 & $<0.018$ & 3\\

HD\,196982 & AT~Mic & MIPS & 24 & $0.1160\pm0.00464$ & 4\\

 &  & MIPS & 70 & $<0.018$ & 4\\
 
 &  & SCUBA & 850 & $<0.009$ & 7\\

HD\,197481 & AU~Mic & MIPS & 24 & $0.1430\pm0.00572$ & 4\\

 &  & MIPS & 70 & $0.2050\pm0.0205$ & 4\\

 &  & MIPS & 160 & $0.1680\pm0.0202$ & 4\\
 
 &  & IRAS & 60 & $0.252\pm0.043$ & 1\\
 
 &  & SHARC & 350 & $0.072\pm0.020$ & 13, 15\\
 
 &  & SCUBA & 450 & $0.085\pm0.042$ & 7\\
 
 &  & SCUBA & 850 & $0.0144\pm0.0018$ & 7\\

HD\,199143 &  & MIPS & 24 & $0.0350\pm0.00014$ & 4\\

 &  & MIPS & 70 & $<0.022$ & 4\\

 &  & MIPS & 160 & $<0.031$ & 4\\
 
 &  & SCUBA & 850 & $<0.0075$ & 7\\

HD\,358623 & AZ~Cap & MIPS & 24 & $0.0130\pm0.00052$ & 4\\

 &  & MIPS & 70 & $<0.0012$ & 4\\
 
 &  & SCUBA & 850 & $<0.0069$ & 7\\
\hline
\multicolumn{6}{l}{\scriptsize{REFERENCES.--(1) \citet{Moor2006}; (2) \citet{Williams2006}; (3) This paper; (4) \citet{Rebull2008}; }} \\ 
\multicolumn{6}{l}{\scriptsize{(5) \citet{Spangler2001}; (6) \citet{Silverstone2000}; (7) \citet{Liu2004}; (8) \citet{Habing2001}; }} \\
\multicolumn{6}{l}{\scriptsize{(9) \citet{Holland1998}; (10) \citet{Liseau2003}; (11) \citet{Moshir1993}; (12) \citet{Hillenbrand2008a} }} \\
\multicolumn{6}{l}{\scriptsize{(13) \citet{Chen2005}; (14) \citet{diFrancesco2008}; (15) \citet{Augereau2006}; (16) \citet{Chini1991} }} \\
\end{longtable}
}
\end{appendix}

\end{document}